\documentclass[pageno,nolineno]{jpaper}

\usepackage{libertinus}
\usepackage{inconsolata}
\usepackage{pmboxdraw}
\usepackage{setspace}

\usepackage[numbers]{natbib}
\usepackage{stmaryrd}
\usepackage{amsfonts}
\usepackage{amsmath}

\usepackage{caption}
\usepackage{subcaption}
\usepackage{mathpartir}
\usepackage{mathtools}
\usepackage{amsmath}
\usepackage{amsthm}

\usepackage{bm}
\usepackage{multirow}
\usepackage[noabbrev,capitalise]{cleveref}
\usepackage{syntax}
\usepackage{enumitem}
\usepackage{cleveref}
\usepackage{xspace}

\usepackage{tikz}
\usetikzlibrary{positioning,shapes,trees,arrows}
\usetikzlibrary{arrows.meta}

\usepackage{float}
\newfloat{listing}{htbp}{lop}
\floatname{listing}{Listing}

\usepackage{array}
\usepackage[framemethod=tikz]{mdframed}
\usepackage{listings-cuda}
\usepackage{listings-descend}
\usepackage{authblk}

\DeclareMathAlphabet{\mathpzc}{OT1}{pzc}{m}{it}

\newcommand{\code}[1]{\mathtt{#1}}
\newcommand{\set}[1]{\{~#1~\}}

\newcommand{\Shrd}{\code{shrd}}
\newcommand{\Uniq}{\code{uniq}}

\newcommand{\Mem}{\code{mem}}

\newcommand{\Nat}{\code{nat}}

\newcommand{\CpuMem}{\code{cpu.mem}}

\newcommand{\GpuGlobal}{\code{gpu.global}}
\newcommand{\GpuShared}{\code{gpu.shared}}

\newcommand{\foreachloop}[3]{\code{for}~#1~\code{in}~#2~\set{#3}}
\newcommand{\fornat}[3]{\code{for}~#1~\code{in}~#2 ~\set{#3}}

\newcommand{\indep}[7]{\code{split}(#1)~#3~\code{at}~#2~\code{\{} #4~\code{=>}~#5\code{,}~#6~\code{=>}~#7\code{\}}}
\newcommand{\sched}[4]{\code{sched(}#1\code{)}~#2~\code{in}~#3~\code{\{}~#4~\code{\}}}

\newcommand{\app}[3]{#1\code{::<}#2\code{>}\code{(}#3\code{)}}

\newcommand{\loan}[2]{\prescript{#1}{}{#2}}

\newcommand{\DataTy}{\delta}

\newcommand{\arrty}[2]{\mathsf{[}#1\mathsf{;}~#2\mathsf{]}}

\newcommand{\gridty}[2]{\code{gpu.Grid}~#1~#2}
\newcommand{\blockty}[1]{\code{gpu.Block}~#1}

\newcommand{\GpuThreadTy}{\code{gpu.Thread}}
\newcommand{\CpuThreadTy}{\code{cpu.Thread}}

\newcommand{\UnitTy}{\code{unit}}

\newcommand{\TupleTy}{{(\DataTy_1, \ldots, \DataTy_n)}}
\newcommand{\ArrayTy}{\arrty{\DataTy}{\eta}}

\newcommand{\Sched}{\sched{\optional{X \opt Y \opt Z}}{x}{e}{t}}

\newcommand{\opt}{~\textcolor{gray}{|}~}
\newcommand{\optional}[1]{\textcolor{gray}{[}#1\textcolor{gray}{]}}

\newcommand{\sdef}{\textcolor{gray}{\Coloneqq}}

\newcommand{\tentry}[4]{$ #1 $ & $ #2 $ & $ #3 $ & \color{gray}#4\\}

\newcommand{\Descend}{\textsf{\textit{Descend}}\xspace}

\title{\Descend: A Safe GPU Systems Programming Language}

\author[1]{Bastian Köpcke}

\author[2]{Sergei Gorlatch}
\affil[1,2]{University of Münster}

\author[3]{Michel Steuwer}
\affil[3]{The University of Edinburgh}

\begin{document}

\date{}
\maketitle

\thispagestyle{empty}

\begin{abstract}
Graphics Processing Units (GPU) offer tremendous computational power by following a throughput oriented computing paradigm where many thousand computational units operate in parallel.
Programming this massively parallel hardware is challenging.
Programmers must correctly and efficiently coordinate thousands of threads and their accesses to various shared memory spaces.
Existing mainstream GPU programming languages, such as CUDA and OpenCL, are based on C/C++ inheriting their fundamentally unsafe ways to access memory via raw pointers.
This facilitates easy to make, but hard to detect bugs such as \emph{data races} and \emph{deadlocks}.

In this paper, we present \Descend: a safe GPU systems programming language.
In the spirit of Rust, \Descend{}'s type system enforces safe CPU and GPU memory management by tracking \emph{Ownership} and \emph{Lifetimes}.
\Descend introduces a new \emph{holistic GPU programming} model where computations are hierarchically scheduled over the GPU's \emph{execution resources}: grid, blocks, and threads.
\Descend's extended \emph{Borrow} checking ensures that execution resources safely access memory regions without introducing data races.
For this, we introduced \emph{views} describing safe parallel access patterns of memory regions.

We discuss the memory safety guarantees offered by \Descend{}'s type system and evaluate our implementation of \Descend{} using a number of benchmarks, showing that no significant runtime overhead is introduced compared to manually written CUDA programs lacking \Descend{}'s safety guarantees.
\end{abstract}

\section{Introduction}
\label{sec:intro}
Graphics Processing Units (GPUs) are massively parallel hardware devices with a throughput oriented design that prioritises the runtime of the overall computation performed in parallel by thousands of collaborating threads over the single thread performance, as classical CPUs do~\cite{DBLP:journals/cacm/GarlandK10}.
This has made GPUs attractive devices in many domains where high performance is crucial, such as in scientific simulations, medical imaging, and most prominently, machine learning.

Writing functionally correct and efficient software for GPUs is a challenging task even for advanced programmers.
The predominant GPU programming languages, CUDA and OpenCL, are low-level imperative systems programming languages, giving programmers great control to precisely influence how each thread accesses memory and when it performs which computational instructions.
This control is needed to extract the expected high performance from GPUs, where the difference between an unoptimized naive implementation and a fully optimized implementation can be up to two orders of magnitude~\cite{10.1145/3570638}--- often significantly more than on CPUs.

Unfortunately, in CUDA and OpenCL, this level of control comes with significant challenges for GPU programmers.
As both languages are based on C/C++ they inherit their fundamentally unsafe ways to access memory via raw pointers.
Furthermore, to coordinate threads and ensure a consistent view of the memory, manual synchronization primitives must be used correctly.
This leads to easy-to-make, but often hard to detect bugs, particularly race conditions when accessing the same memory location from multiple threads and deadlocks when using the synchronization primitives incorrectly.

\Cref{lst:cuda-transpose-example} shows a CUDA kernel function, executed in parallel on the GPU to transpose a matrix.
In lines 4--7, each thread copies four matrix elements into a temporary buffer and then---after a synchronization---copies the transposed elements to the output.
The correctness of this function depends on correct indexing which is notoriously tricky.
In fact, \cref{lst:cuda-transpose-example} contains a subtle bug:
In line 5, \lstinline!threadIdx.y+j! should be enclosed by parenthesis, so that both terms are multiplied by 32.
As a result, a data race occurs as multiple threads will write uncoordinated into the same memory location.

\begin{listing}[b]
  \begin{CUDAListing}
__global__ void transpose(const double *input,
                                double *output) {
  __shared__ float tmp[1024];
  for (int j = 0; j < 32; j += 8)
     tmp[ threadIdx.y+j *32+threadIdx.x] =
        input[(blockIdx.y*32+threadIdx.y+j)*2048
                        + blockIdx.x*32+threadIdx.x];
  __syncthreads();
  for (int j = 0; j < 32; j += 8)
     output[(blockIdx.x*32+threadIdx.y+j)*2048
                      + blockIdx.y*32+threadIdx.x] =
        tmp[(threadIdx.x)*32+threadIdx.y+j];    }
  \end{CUDAListing}
  \vspace{-0.5cm}
  \caption{A CUDA GPU kernel performing a matrix transposition in parallel. A subtle bug in the indexing in line 5 leads to a data race.}
  \label{lst:cuda-transpose-example}
  \end{listing}

Rust has demonstrated that a systems programming language can be designed in a memory safe way without losing low-level control.
It prevents data races, by forbidding the concurrent access of threads to a memory resource if at least one thread is allowed to mutate it~\cite{DBLP:journals/cacm/JungJKD21}.
Rust enforces this with its type system, specifically with \emph{borrow} checking, that interacts with the concepts of \emph{ownership} and \emph{lifetimes} which primarily ensure safe memory management.
Could Rust have prevented the bug in \cref{lst:cuda-transpose-example}?
Clearly, \lstinline!tmp! is shared among the parallel executing threads and, clearly, we mutate it's content in line 5.
Therefore, Rust would reject this kernel, even without attempting to investigate if the indexing is safe, as Rust's type system has no capabilities of reasoning about safely accessing an array in parallel by multiple threads.

In this paper, we introduce \Descend{}, a safe GPU systems programming language adapting and extending the ideas of Rust towards GPU systems.
In contrast, to prior safe GPU programming solutions, such as Nvidia's Thrust~\cite{thrust} or Futhark~\cite{DBLP:conf/pldi/HenriksenSEHO17}, \Descend{} is an imperative GPU programming language empowering the programmer to exercise low-level control with a safety net.

\Cref{lst:descend-transpose-example} shows the matrix transposition function in \Descend{}.
In contrast to CUDA, this function is not implicitly executed by thousands of GPU threads, instead this function is executed by the (one) GPU grid.
Programmers describe the hierarchical scheduling of the computation over the \emph{grid}, first by describing the scheduling of \emph{blocks} (line~4) and then the nested \emph{threads} (line~6).
For each \lstinline!block! we allocate shared memory in line~5.
Each \lstinline!thread! performs the same copies as in CUDA, first from the input into the temporary buffer, and then---after a synchronization--- back into the output.
Instead of raw indexing, in \Descend{} programmers use memory \emph{views} to describe parallel accesses into memory.
\Descend{} statically checks that accesses into views are safe, and treats them specially in the type system.
This restricts memory accesses to safe parallel access patterns, while still allowing compositions of views for describing complex memory accesses.
For the example, the borrow checking of \Descend{} is capable to statically determine that the parallel write access into the shared temporary buffer and the output are safe.
Similarly, \Descend{} statically enforces the correct use of the synchronization, that can not be forgotten or placed incorrectly.

\begin{listing}[t]
\begin{DescendListing}
fn transpose(input:  &     gpu.global [[f64;2048];2048],
             output: &uniq gpu.global [[f64;2048];2048])
 -[grid: gpu.grid<XY<64,64>,XY<32,8>>]-> () {
  sched(Y,X) block in grid {
    let tmp = alloc::<gpu.shared, [[f64; 32]; 32]>();
    sched(Y,X) thread in block {
      for i in [0..4] {
        tmp.group_by_row::<32,4>[[thread]][i] =
          input.group_by_tile::<32,32>.transpose[[block]]
                .group_by_row::<32,4>[[thread]][i] };
      sync;
      for i in [0..4] {
        output.group_by_tile::<32,32>[[block]]
              .group_by_row::<32,4>[[thread]][i] =
                  tmp.group_by_row::<32,4>[[thread]][i] }
    } } }
\end{DescendListing}
\vspace{-0.5cm}
\caption{A \Descend{} function performing matrix transposition.}
\label{lst:descend-transpose-example}
\end{listing}

\Descend{} is a holistic programming language for heterogeneous systems comprised of CPU and GPU.
The physically separated memories of CPU and GPU are reflected in the types of references for which \Descend{} enforces that they are only dereferenced in the correct execution context.
Functions are annotated with an \emph{execution resource} (as seen in the function signature in line~3) that indicates how a function is executed.
These annotations make important assumptions, that are implicit in CUDA, about how many threads and blocks execute a kernel, explicit and enforcable by the type system.

In summary, this paper makes the following contributions:
\begin{itemize}
  \item we introduce \Descend{}, a safe GPU systems programming language in the spirit of Rust;
  \item we identify the challenges of GPU programming and discuss how \Descend{} assists in  addressing them (\Cref{sec:gpu-programming-difficult});
  \item we discuss how the concepts of \emph{execution resources}, \emph{place expressions}, and memory \emph{views} ensure safety (\Cref{sec:descend});
  \item we present \Descend{}'s formal type system and extended borrow checking (\Cref{sec:descend-types});
  \item and show in an experimental evaluation that programs written in \Descend{} achieve the same performance as equivalent programs written in CUDA, that lack \Descend{}'s safety guarantees (\Cref{sec:eval}).
\end{itemize}
We discuss related work and conclude in sections \ref{sec:relatedWork} and \ref{sec:conclusion}.

\section{Challenges of GPU Programming}
\label{sec:gpu-programming-difficult}
GPU programming brings a number of challenges, that we discuss in this section.
We group them in two areas:
\emph{1)} challenges that come from working with the execution and memory hierarchies of GPUs, such as thousands of threads grouped in blocks accessing various GPU memories; and
\emph{2)} challenges that come from coordinating the heterogeneous system, such as transferring data between CPU and GPU memory.
Before we discuss each area, we give a brief overview of the traditional GPU programming model established by CUDA.

\subsection{The CUDA GPU Programming Model}
In CUDA, programmers write \emph{kernel} functions that are executed in parallel on the GPU, often performing data parallel computations over multidimensional arrays.
These functions are executed by many thousand threads, all executing the same code.
Therefore, on a first view, the CUDA programming model resembles traditional data-parallel programming models, where a single instruction is applied to multiple data elements in lock-step.
However, in CUDA this strict requirement is relaxed as the kernel code can branch based on the \emph{thread index}.
The \emph{thread index}, identifies the individual thread and is usually used for indexing into arrays so that each thread processes a different array element.
Thread indices are plain integers involved in index computations into plain C-style arrays, making statically checking the safety of parallel memory accesses challenging and leading to data races being introduced by easy to make bugs.
Furthermore, kernels are often written with implicit assumptions about how many threads execute them, making them hard to understand without knowing these assumptions and an additional source of bugs, when CPU and GPU code start to diverge.

GPUs are comprised of multiple multithreaded Streaming Multiprocessors (SM), each capable of executing multiple threads simultaneously.
It makes sense to reflect this hardware design in the software.
Therefore, \emph{threads} are hierarchically organized into groups, that are executed independently by the SMs.
In CUDA, such groups of threads are called \emph{blocks}.
The collection of all blocks is called the \emph{grid}.

Similarly, memory is organized hierarchically as well and closely connected to the execution hierarchy.
In software, separate \emph{address spaces} reflect the different kinds of GPU memory.
The slowest and largest memory is \emph{global memory}, which is accessible by each thread in the entire grid.
Each block provides the fast \emph{shared memory} which is accessible only by each thread in the block.
Lastly, each thread has exclusive access to its own and fastest \emph{private memory}.
Data transferred from the host to the GPU is always stored in global memory.
In order to exploit the faster memories, data has to be explicitly copied between address spaces.

\subsection{Challenges of the Execution \& Memory Hierarchies}
The CUDA programming model with its execution and memory hierarchies, resembles closely the GPU hardware and enables scalability of GPU programs, but it comes with two major challenges:
how to avoid \emph{data races} and how to correctly \emph{synchronize} the threads of a block.

\paragraph{Data Races.}
Data races occur when two or more threads access the same memory location at the same time and at least one thread writes to the memory location.
It is very easy to create a data race in CUDA, consider the following code:
\begin{CUDAListing}
__global__ void rev_per_block(double *array) {
  double *block_part = &array[blockIdx.x * blockDim.x];
  block_part[threadIdx.x] =
        block_part[blockDim.x-1 - threadIdx.x]; }
\end{CUDAListing}
In this example, the input \lstinline!array! is split into independent parts for each block.
Then the threads in each block access a single element in the reverse order of their thread index and write the value back into the array at their thread index.
This creates a data race:
a thread may still be reading a value from an index that another thread is already writing to.

In \Descend{}, the compiler recognizes the possibility of a data race and would reject the program with an error message:
\begin{DescendListing}
error: conflicting memory access
  | arr[[thread]] = arr.rev[[thread]];
  |                 ^^^^^^^^^^^^^^^^^-------------------
  | ^^^^^^^^^^^^^ cannot select memory because of       |
  |               a conflicting prior selection here: --
\end{DescendListing}
We will explain in \cref{sec:descend}, that for this check \Descend{} performs an extended \emph{borrow (or access) checking} similar to Rust, tracing which memory location (formalized as \emph{place expressions}) is accessed by which thread (formalized as \emph{execution resources}).
To make this check feasible, in \Descend{} programmers express parallel memory accesses via \emph{views}, which are safe parallel access patterns, such as \lstinline!rev! for reverse in this example.
Views can be composed to enable complex parallel access patterns that are still known to be safe.

\paragraph{Synchronization.}
Because blocks run independently, all blocks are only synchronized when a GPU kernel finishes execution.
However, as the threads of a block are executed on the same SM it is possible to synchronize them using a block-wide barrier.
To avoid undefined behavior, including possibly a deadlock, every thread in the block must reach the barrier.
Unfortunatly, is it easy to violate this requirement:
\begin{CUDAListing}
__global__ kernel(...) {
  if (threadIdx.x < 32) { __synchronize() } }
\end{CUDAListing}
In this CUDA kernel, the \lstinline{__synchronize} barrier is executed only by threads that have an index smaller than $32$ within each block.
When launched with up to $32$ threads per block, each thread reaches the barrier as required, but when launched with more than $32$ threads per block, the behaviour of the program is undefined.

In \Descend{}, a program such as this would not compile, if there are more than $32$ threads per block.
The equivalent \Descend{} program would fail with an error message like this:
\begin{DescendListing}
error: barrier not allowed here
  |  split(X) block at 32 {
  |  ^^^^^^^^^^^^^^^^^ `block` is $\texttt{split}$ here --------
  |    first_32_threads => { sync }                 |
  |                          ^^^^ `$\texttt{sync}$` not        |
  |      performed by all threads $\texttt{in}$ the block ------
\end{DescendListing}

We will discuss in \Cref{sec:descend}, how \Descend{} checks that synchronizations are performed correctly.
In \Descend{}, either all threads in a block perform the same instructions (when using the \lstinline[language=descend]!sched! syntax seen before), or the threads in a block must be split using the syntax shown in the error message above.
Basically, a synchronization nested inside of a split block is forbidden.
\Descend{} also ensures that synchronizations are not forgotten.
A synchronizations releases borrows of the synchronized memories which, if forgotten, are flagged by the borrow checker as seen above.

\subsection{Challenges of Heterogeneity}
GPUs are not programmed only by writing kernels.
They are part of a heterogeneous system with CPU and GPU performing computations asynchronously and the CPU managing the computations on the GPU.
Two significant challenges arise from this:
the handling of the physically \emph{separated memories} on CPU and GPU,
and dealing with \emph{shared assumptions} between CPU and GPU that are often not explicitly encoded and can break correctness in subtle ways.

\paragraph{Separated Memories.}{
The CPU and GPU are physically distinct devices.
A \emph{host} thread running on the CPU launching GPU kernel and transferring data to and from the GPU via an API without direct access to the GPU's memory.
The CPU program only accesses CPU memory, while a GPU program only accesses its various GPU memories.
CUDA reflects the separated memories via \emph{address spaces} that are annotations on pointers.
These annotations are not strickly enforced by the CUDA compiler, making it easy for programmers to make mistakes that are not caught by the compiler, such as misusing the provided API for copying data to the GPU:
\begin{CUDAListing}
cudaMemcpy(d_vec, h_vec, size, cudaMemcpyDeviceToHost);
\end{CUDAListing}
Function \lstinline{cudaMemcpy} copies size many bytes to the destination in the first argument from the source in the second argument.
The last argument specifies whether the destination and source are on the device or host.
In the above call, destination and source pointers are swapped, which leads to the address in the host pointer being used to access memory on the device, with undefined behavior.

In \Descend{}, reference types carry additional information and correct usage is strictly enforced.
Making the same mistake as above, leads to an error message at compile time:
\begin{DescendListing}
error: mismatched types
  | copy_mem_to_host(d_vec, h_vec);
  |                  ^^^^^ expected reference to
  |      `gpu.global`, found reference to `cpu.mem`
\end{DescendListing}

In CUDA, CPU pointers can be directly passed to the GPU.
This means, that a GPU program may accidentally attempt to directly access CPU memory, as in the following code:
\begin{CUDAListing}
void host_fun() {
  double *vec = malloc(sizeof(double) * N * N);
  init_kernel<<<N, N>>>(vec);  }

__global__ void init_kernel(double *vec) {
  vec[globalIdx.x] = 1.0; }
\end{CUDAListing}
In this example, the host allocates space for an array in the CPU main memory and passes the resulting pointer to the GPU.
The GPU program then attempts to initalize the memory, but it has no access to the separated main memory, leading to undefined behavior.

In \Descend{}, a program such as this would not compile, because the compiler recognizes that we are attempting to access CPU memory on the GPU.
The equivalent GPU program in Descend would fail like this:
\begin{DescendListing}
error: cannot dereference `*vec` pointing to `cpu.mem`
  |  sched(X) thread in grid {
  |  -----------------------     executed by `gpu.Thread'
  |     (*vec)[[thread]] = 1.0
  |      ^^^^ dereferencing pointer $\texttt{in}$ `cpu.mem' memory
\end{DescendListing}
}
In \Cref{sec:descend}, we introduce \emph{execution resources} that identify \emph{who} executes a piece of code with a focus on the GPU.
However, these also extend to CPU threads.
The formal type system, introduced in \Cref{sec:descend-types}, extends references with \emph{memory annotations}  that strickly enforce that memory is only dereferenced in the correct execution context.

\paragraph{Shared Assumptions between CPU and GPU.}{
In CUDA---and \Descend{}---when launching a function on the GPU, the host thread specifies the launch configuration, i.e., the amount of threads executing the kernel and their grouping into blocks.
When implementing GPU functions, there are often implicit assumptions about the amount of threads that are going to execute the function as well as the amount of memory that is allocated via the host's memory API.
But these assumptions can easily be violated on either the CPU or GPU side, such as for this GPU function scaling a vector:
\begin{CUDAListing}
__global__ scale_vec_kernel(double *vec) {
  vec[globalIdx.x] = vec[globalIdx.x] * 3.0; }
\end{CUDAListing}
Each thread on the GPU accesses a single element of the vector at its index within the entire grid.
The assumption made here, is that the grid contains as many threads as there are elements in the vector.
For example, the following launch of the GPU function from the CPU is erroneous:
\begin{CUDAListing}
cudaMalloc(&d_ptr, SIZE);
...
scale_vec_kernel<<<1, SIZE>>>(d_ptr);
\end{CUDAListing}
Instead of starting as many threads as there are vector elements, the function is executed by as many threads as there are \emph{bytes} in the vector.
By launching the GPU function with more threads than vector elements, out-of-bounds memory accesses are triggered.

In \Descend{}, calling a GPU program with the wrong amount of threads leads to an error message at compile time:
\begin{DescendListing}
error: mismatched types
  |  scale_vec<<<X<1>, X<SIZE>>>>(d_vec);
  |                               ^^^^^
  |      expected `[f64; SIZE]`, found `[f64; ELEMS]`
\end{DescendListing}

We will see in \Cref{sec:descend}, that all functions are annotated with an execution resource describing how the function expects to be executed.
This makes assumptions explicit.
The type system, presented in \Cref{sec:descend-types}, enforces this at compile time and supports both launch configurations with constants (such as $1024$) and variables via polymorphism.

\section{Safe GPU programming with \Descend{}}
\label{sec:descend}
\begin{figure*}[t]
  \centering
  \begin{subfigure}[t]{.2\textwidth}
    \centering
    \includegraphics[width=\textwidth]{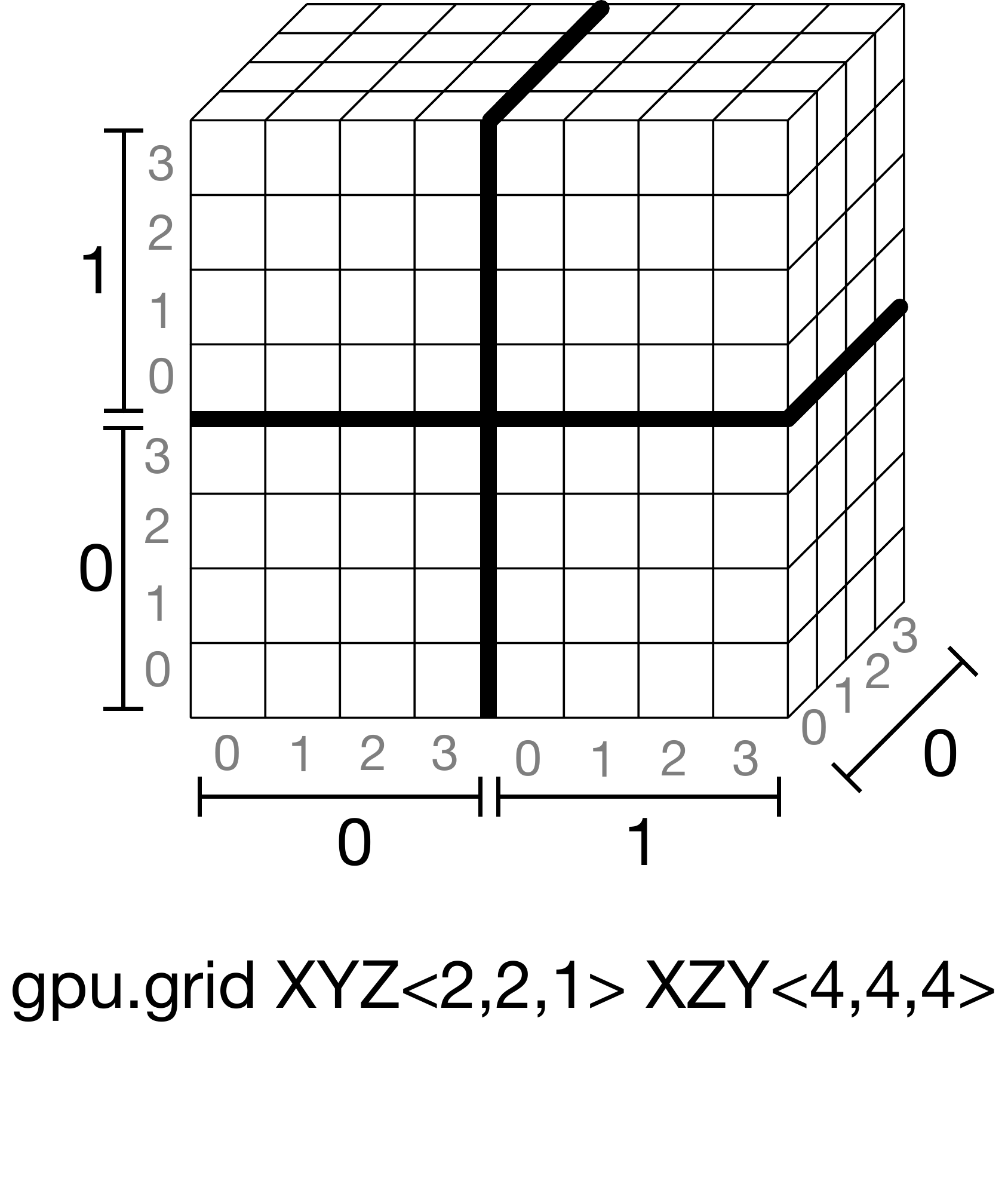}
    \caption{A 3D grid of four blocks each with $4\times{}4\times{}4$ threads.}
    \label{subfig:grid}
  \end{subfigure}
  \hspace{3em}
  \begin{subfigure}[t]{.2\textwidth}
    \centering
    \includegraphics[width=\textwidth]{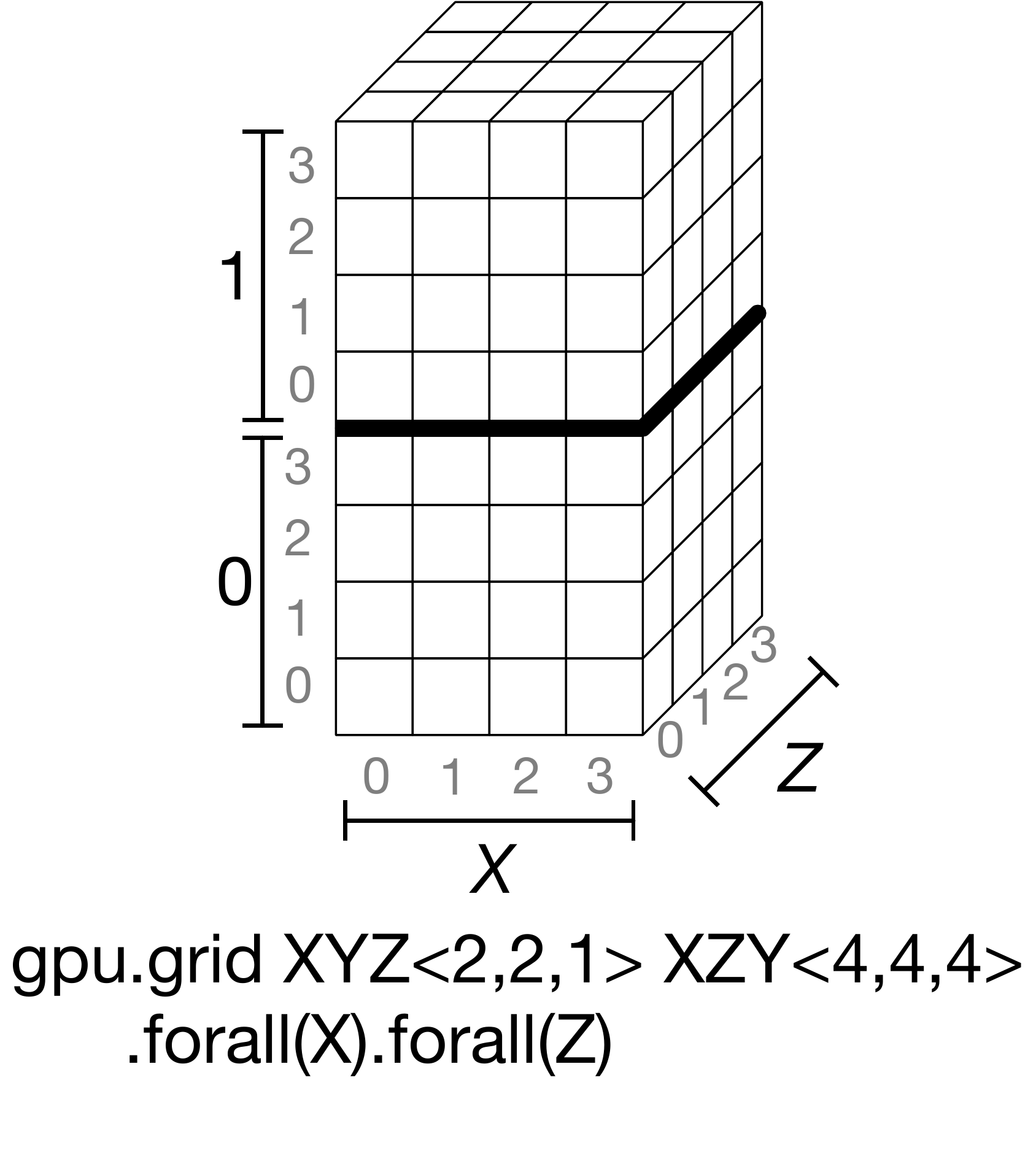}
    \caption{A group of two blocks, after scheduling in $X$ and $Z$ dimension.}
    \label{subfig:sched-grid}
  \end{subfigure}
  \hspace{3em}
  \begin{subfigure}[t]{.2\textwidth}
    \centering
    \includegraphics[width=\textwidth]{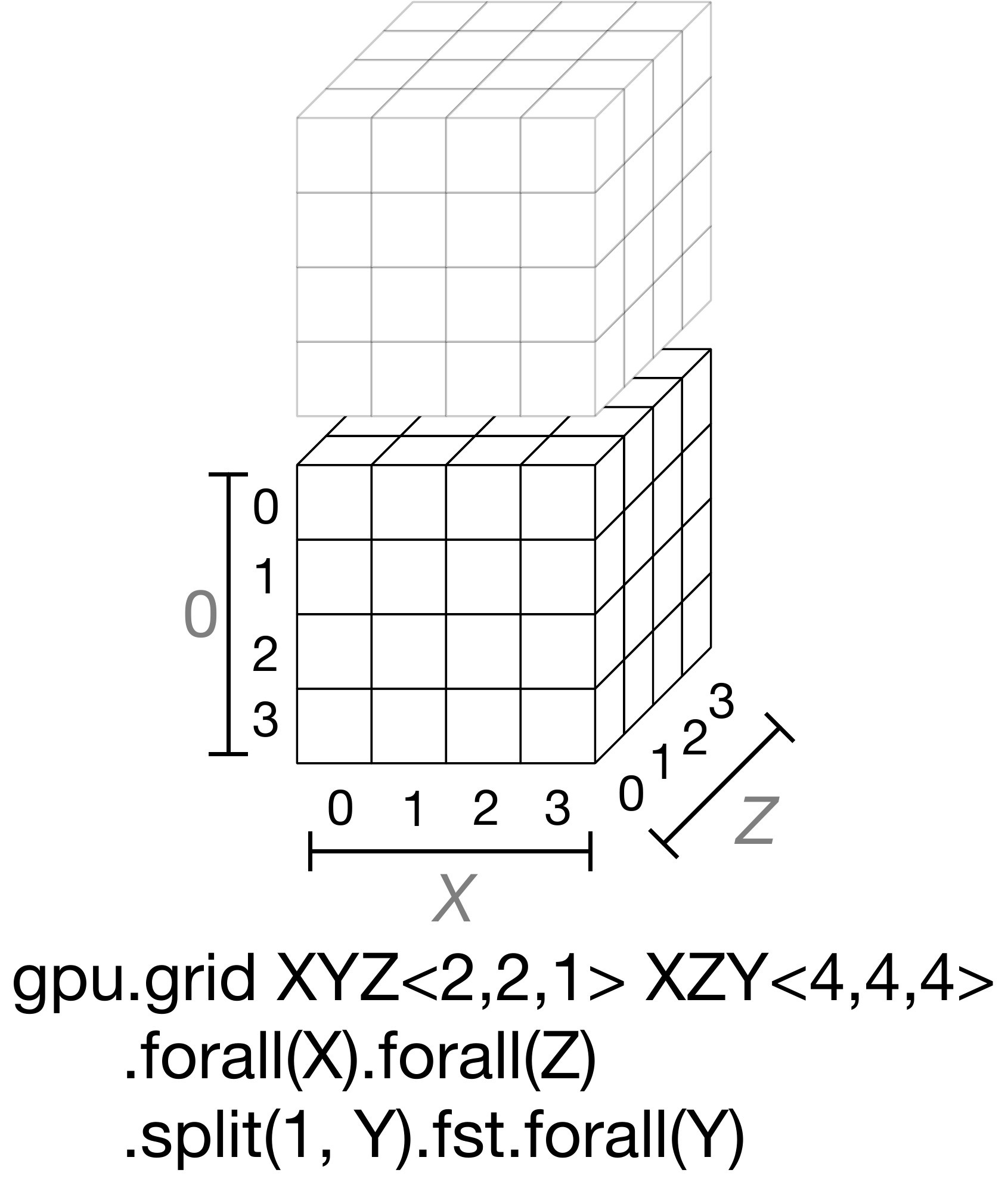}
    \caption{A single block of threads selected using \lstinline!split!.}
    \label{subfig:split-grid}
  \end{subfigure}
  \caption{Visualization of hierarchically scheduling of execution resources in \Descend{}}
\end{figure*}
In this section, we discuss the technical mechanism that \Descend{} uses to guarantee memory safety and produce the error messages seen in the previous section.
We first give explanations and intuitions, before we will present the most important aspects of the formal type system of \Descend{} in \Cref{sec:descend-types}.

We start by introducing \emph{execution resources}, \emph{place expressions}, and \emph{views} as the central ingredients to formally reason about the execution and memory hierarchy, and to check that parallel memory accesses are performed safely.

\subsection{Execution Resources}
In \Descend{} computations on the GPU are hierarchically scheduled over the grid of blocks and threads.
In the following example code, we execute a function with a 3D GPU grid of $2\times{}2\times{}1$ blocks, each comprised of $4\times{}4\times{}4$ threads.
The shape of this grid is visualized in \cref{subfig:grid} and described by an \emph{execution resource}, as indicated in line 1:
\vspace{-1em}
\begin{DescendListing}
fn foo(...) -[grd: gpu.Grid<XYZ<2,2,1>,XYZ<4,4,4>>]-> (){
  sched(X,Z) blocks in grd {
    split(Y) blocks at 1 {
        fstBlock => ...
        sndBlock => ...  } } }
\end{DescendListing}

Here, \lstinline!grd! is the execution resource and we specify it by annotating its type describing the shape of the grid.
The next line is ``executed'' by the entire grid.
We write this in quotation marks, as execution resources besides individual threads are only allowed to either allocate memory, or to schedule computations over their nested execution resources.

We schedule the following computation over all groups of blocks in the grid with the same $X$ and $Z$ coordinates.
Such a group of blocks is visualized in \cref{subfig:sched-grid}.
In the code, the execution resource is \lstinline!blocks! and formally \Descend{} treats it as an alias of
\lstinline!gpu.grid<XYZ<2,2,1>,XZY<4,4,4>>.forall(X).forall(Z)!, which will be used for performing safety checks as discussed later.
We can see, that the execution resource tracks all information about how the grid has been hierarchically scheduled up to this point.

In line 3 of the example, we split the blocks in the $Y$ dimension at position 1 into two subgroups.
Each subgroup is identified by a separate execution resource \lstinline!fstBlock! and \lstinline!sndBlock! which are allowed to perform independent computations.
\cref{subfig:split-grid}, visualizes \lstinline!fstBlock! and shows the formal representation of the execution recourse that \Descend{} reasons with.

Using \lstinline!sched! and \lstinline!split! the grid is scheduled hierarchically, until a single thread  is reached performing computations.

{%
\floatstyle{boxed}
\restylefloat{figure}
\begin{figure}
  \begin{tabular}[t]{ l c l >{\em}r }
    \tentry {e}{ \sdef }
          { \code{cpu.thread} }{Exec.\,Res. }
  \tentry {}{}
          { \code{gpu.grid}\langle{}d,d\rangle{}}{}
  \tentry {}{}
          { e \code{.forall([X \opt Y \opt Z])} }{}
  \tentry {}{}
          { e \code{.split(\eta, [X \opt Y \opt Z]).[fst \opt snd]} }{}
  \tentry { d }{ \sdef }
          { \code{XYZ\langle}\eta, \eta, \eta\code{\rangle} }{ Dims. }
  \tentry {}{}
          {\code{XY\langle}\eta, \eta\code{\rangle} \opt \code{XZ\langle}\eta, \eta\code{\rangle} \opt \code{YZ\langle}\eta, \eta\code{\rangle}}{}
  \tentry {}{}
          {\code{X\langle}\eta\code{\rangle} \opt \code{Y\langle}\eta\code{\rangle} \opt \code{Z\langle}\eta\code{\rangle}}{}
  \tentry{ \eta }{ \sdef }{ 0 \opt \ldots \opt 9 \opt n \opt \eta + \eta \opt \eta * \eta \opt \ldots}{Nats.}
\end{tabular}
  \caption{Execution Resources and Dimensions}
  \label{syn:execs}
\end{figure}
}

\Cref{syn:execs}, shows the formal grammar of execution resources.

Besides the hierarchical GPU grid, we have an execution resource to describe CPU threads (\lstinline!cpu.thread!).
This is used, for example, to mark functions executed on the CPU.
The \lstinline!gpu.grid! stores two \emph{dimensions} $d$ that describe the number and up to three-dimensional shape of the blocks.
We add the additional 1D and 2D forms, such as \lstinline!XY<$\eta$,$\eta$>!, to be able to check that we do not schedule over a missing dimension.
The size of a dimension is represented as a natural number $\eta$ that can either be a constant, a variable, or simple mathematical expressions over natural numbers.

GPU grids or blocks can either be scheduled by treating all of their elements the same using the \lstinline!sched! syntax seen before and represented using the \lstinline!forall! notation in the execution resource indicating which dimension to schedule over.
Alternatively, we might \lstinline!split! an execution resource at a position into two distinct subgroups performing independent instructions, as seen in the example above.
This is represented with the \lstinline!split! notation in the execution resource, where we also immediatly must select one of the two subgroups.
Looking back at the example from the beginning of the section, \lstinline!fstBlock!corresponds to splitting the blocks and then immediately selecting the first block (\lstinline!blocks.split(1, Y).fst!).

The execution resources introduced here have three main purposes in \Descend{}:
\emph{1)} they are used to check what code is executed on the CPU and GPU;
\emph{2)} they are used to check what instructions are executed by which part of the GPU hierarchy, such as that a barrier synchronisation must be executed inside a block;
\emph{3)} they keep track of dimensions and sizes used in the code generation.

\subsection{Place Expressions and Views}
\label{sec:place-expressions}
{%
\floatstyle{boxed}
\restylefloat{figure}
\begin{figure}
\begin{tabular}[t]{ l c l >{\em}r }
\tentry{ p }{ \sdef }{}{ Place Expressions: }
\tentry{}{}{ x }{ variable }
\tentry{}{}{p.fst \opt p.snd}{ projections }
\tentry{}{}{*p}{ dereference }
\tentry{}{}{p[t]}{ index }
\tentry{}{}{p\llbracket e \rrbracket}{ select }
\tentry{}{}{p .v\code{::\langle \overline{\eta}, \overline{\delta}\rangle(}v\code{)}}{ views }
\end{tabular}
  \caption{Place Expressions}
  \label{syn:pl-expr}
\end{figure}
}

\paragraph{Place Expressions}
Rust introduces the concept of a \emph{place expression} as unique names for a region of memory.
Aliases are resolved by substituting the referenced place expressions.
This allows them to be compared syntactically in Rust's type system to make sure the same memory location is not (mutably) accessed at the same time.
Through this, it can be guaranteed that no data races occur.

\Cref{syn:pl-expr} shows the place expressions that exist in \Descend{}.
The simplest place expression is a variable, which names a region of memory.
Projections $\code{.fst}$ or $\code{.snd}$ are applied to tuples referring to two non-overlapping regions of memory.
The dereference-operator accesses the memory that a reference points to.
Single elements of a region of memory containing an array of data of the same type, are accessed by indexing.
All of these place expressions exist in Rust as well.

In \Descend, we introduce two additional place expressions.
The select expression selects memory for an execution resource from an array.
This operation requires the execution resource (such as a block) to consists of as many independent sub-execution resources (such as threads) as there are elements in the array.
Then, each sub-execution resource accesses one element of the array, providing a safe concurrent array access.
However, this access is very restricted.
There is no means of selecting multiple elements yet.
Neither can we change which elements are accessed by which sub-execution recourse.
To increase the flexibility of safe parallel memory accesses we introduce \emph{views}.

\paragraph{Views}

A view reshapes an array or reorders its elements and, therefore, transforms the way we access the array.
Views are applied to place expressions that refer to arrays.
Applying a view to a place expression results in a new place expression, which allows for chaining multiple views.
The memory layout of the original array stays the same.
Only the behavior when accessing the array changes according to the view.
When generating code, views are compiled into raw indices following a process similar to the one taken in the Lift compiler~\cite{DBLP:conf/cgo/SteuwerRD17} and DPIA~\cite{DBLP:journals/corr/abs-1710-08332}.

\Cref{lst:view-transforms} shows the types of the basic views in \Descend{}.
$\code{split}$ splits the array into two non-overlapping partial arrays at a position and returns a tuple containing them.
The split position must be within the size of the input array.
This view allows programmers to select only a part of an array to work on and do something else with the other part or discard it.
View $\code{group}$ combines consecutive elements in the input array into nested arrays of a given size.
Therefore, by grouping elements, the dimensionality of the array is increased by one.
The nested arrays form the elements of the outer array.
This enables selecting entire groups of elements with the select operator.
View $\code{transpose}$ transposes a two dimensional array and $\code{reverse}$ reverses the order of elements.
Finally, $\code{map}$ applies a view to each element of an array.

We compose basic views to form new more complex views.
This is, for example, the case for the views used in \cref{lst:descend-transpose-example}.
Specifically, view \lstinline{group_by_row} is defined as
\vspace{-1em}
\begin{DescendListing}
view group_by_row<row_size: nat, num_rows: nat> =
  group::<row_size/num_rows>.map(transpose)
\end{DescendListing}

\begin{listing}[t]
\begin{DescendListing}
split<k: nat, n: nat, d: dty>([[d; n]])
  -> ([[d; k]], [[d; n-k]]) where n >= k
group<k: nat, n: nat, d: dty>([[d; n]])
  -> [[ [[d; k]]; n/k]] where n 
transpose<m: nat, n: nat, d: dty>([[ [[d; n]]; m]])
  -> [[ [[d; m]]; n]]
reverse<n: nat, d: dty>([[d; n]]) -> [[d; n]]
map<m: nat, n: nat, d1: dty, d2: dty>(
    ([[d1; m]]) -> [[d2; m]], [[ [[d1; m]]; n]] )
  -> [[ [[d2; m]]; n]]
\end{DescendListing}
\caption{Basic views with their types.}
\label{lst:view-transforms}
\end{listing}

\begin{figure}
  \centering
  \includegraphics[width=.48\textwidth]{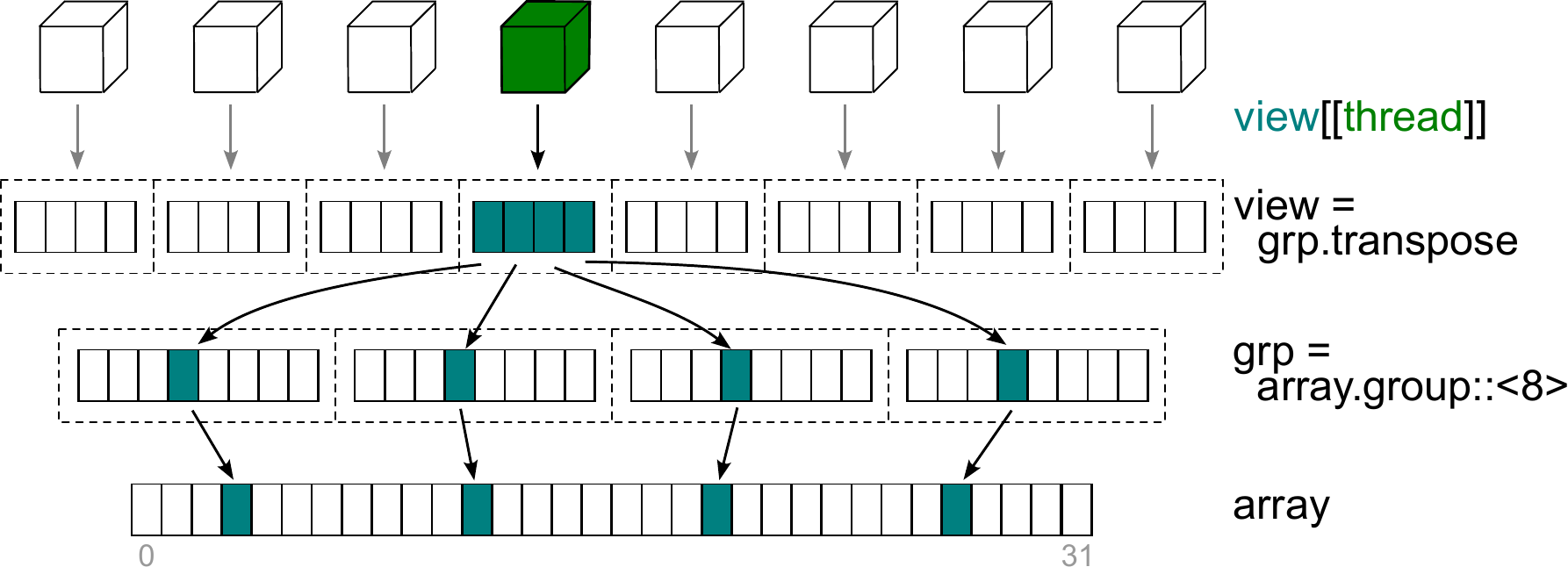}
  \caption{Safe parallel access by thread at the top to highlighted elements of an array at the bottom via views.}
  \label{fig:select-view}
\end{figure}

\Cref{fig:select-view}, shows a group of threads, at the top, safely accessing an array, at the bottom.
The full place expression \lstinline!array.group::<8>.transpose! describes transforming our view of the array.
We first group the array of 32 elements into 4 groups of 8 elements each and call the resulting two-dimensional array \lstinline{grp}.
Then we transpose the grouped array, before we access it in parallel with multiple threads using the select operator.
This array access is safe by construction, as each view describes a safe reshaping of the array resulting only in a remapping of which thread accesses which array elements.

For checking that a place expression, is accessed exclusively, \Descend{}, like Rust, compares the differences between place expressions syntactically.
For example, we can syntactically determine that the place expressions \lstinline!x.split::<32>.fst! and \lstinline!x.split::<32>.snd! are distinct (as they are split non-overlappingly at the same position), while both overlap with the place expression \lstinline!x!, representing the entire array.

\subsection{Extended borrow checking in \Descend{}}

\paragraph{What is Rust's borrow checker and how does it work?}
Rust introduces the concepts of \emph{ownership}, \emph{borrowing}, and \emph{lifetimes} to statically guarantee that there are no data races, memory leaks, use-after-free errors and other memory related errors in the program.
Ownership ensures that there is no aliasing of memory objects, as on assignment to a new variable the value can only be accessed via the new variable.
Attempts to access the old variable lead to compiler errors.

As this model it too restrictive, with borrowing, a restricted notion of aliasing is introduced into the language.
The \emph{borrow checker} checks if a thread is allowed to create a reference to, i.e., ``borrow'', a memory object.
References are either unique or shared.
Multiple shared references can be used at the same time, but only for reading and not writing.
A unique reference, guarantees that there are no other references or variables that can be used to access the same memory location.
It is therefore safe to mutate the underlying memory.

Finally, lifetimes ensure that the memory a reference refers to still exists and hasn't been deallocated.
Attempting to dereference at a point in the program at which the underlying memory has been freed results in a compiler error.

\paragraph{\Descend{}'s extended borrow checker}
On the CPU, \Descend{} implements exactly the same rules as Rust.
On the GPU side, the ownership and borrowing rules are extended and diverge from Rust.
In Rust, exclusive ownership always belongs to a single thread.
In \Descend{}, each execution resource, such as the grid or a block might take ownership of a memory object.
Analogously, execution resources might create references, i.e., they might borrow.
This means that collections of blocks or threads, as well as single threads, own and borrow memory objects, formally represented as place expressions.
The parameters of a function, are owned by the function's execution resource.
In order for a single thread to be able to write into a memory location by having exclusive access to it, the ownership and borrows must be \emph{narrowed} using \Descend{}'s hierarchical scheduling, selections and views.

\textbf{Narrowing}
Narrowing describes how ownership and borrows are refined when navigating the execution hierarchy from grid, to blocks and threads.
For example, the ownership of an array by a grid is narrowed to the grid's blocks by the blocks collectively borrowing the array, each block a distinct part.
This might be further narrowed to the block's threads.

But narrowing can also be violated, as shown here:
\begin{DescendListing}
fn kernel(arr: &uniq gpu.global [f32; 1024])
          -[grd: gpu.Grid<X<32>,X<32>>]-> () {
  sched(X) block in grd {
    let in_borrow = &uniq *arr; // Narrowing violated
    sched(X) thread in block {
      let grp = &uniq arr.group::<32>[[thread]];
                                // Narrowing violated
      arr.group::<32>[[block]][[thread]]; } } }
\end{DescendListing}

In the example, the parameter \lstinline{arr} is owned by the grid.
Attempting to borrow \lstinline{arr} in line 4 \emph{after} having scheduled the blocks of the grid violates narrowing, because each block in the grid would get unique writing access to the entire array.

Another narrowing violation, is shown in line 6.
Here the array is grouped so that there are as many groups as there are threads per block.
Then each thread selects a group and borrows that group uniquely.
However, the selection is performed for each block, as no selection for the \lstinline!block! has been performed.
Therefore, threads from different blocks would gain access to the same memory location.

Line 8 shows correct narrowing.
The array is grouped and each block exclusively borrows a part of the array, before each thread in each block selects an element from it.

\paragraph{Synchronization}
\Descend's system of narrowing ensures that no two threads have mutable access to the same memory location.
However, sometimes we do want to communicate with another thread via shared memory and then the other thread must be able to access the same memory location as well.
We, therefore, need a way to allow a subsequent access by another thread while guaranteeing that this access cannot lead to a data race.
By synchronizing threads at a barrier, we get the guarantee that all memory accesses before the barrier cannot conflict with memory accesses after the barrier.

\medskip
In this section we discussed how \Descend addresses two challenges identified in \cref{sec:gpu-programming-difficult}: preventing data races and handling synchronizations.
We will see in \Cref{sec:descend-types} that in the type system, \Descend tracks in an  environment of the typing judgement a mapping of which execution resource accessed which place expressions.
This environment is used in the typing rules to perform the access safety checks, including correct narrowing.
On a synchronization, we remove from the mapping the previous accesses of all threads in the block.
In the next sections, we discuss how \Descend{} addresses the challenges of managing the heterogeneous system.

\subsection{Handling Separated Memories in \Descend{}}
\label{subsubsec:mem-and-borrow}

\paragraph{Tracking Memory Spaces}

\Descend annotates reference types with \emph{address spaces}.
This is similarly done in CUDA, but CUDA does not have an address space for CPU pointers and generally does not strictly enforce their correct use.
In \Descend, all references carry an address space, including the $\CpuMem$ address space that comprises values stored in the CPU stack and heap.
For the GPU, we differentiate between the global $\GpuGlobal$ and shared memories $\GpuShared$, which have separate address spaces.
Using execution resources, \Descend enforces that references are only dereferenced in the correct execution context, such as preventing dereferencing a GPU reference on the CPU.

\Descend also supports polymorphism over memory spaces, by introducing a type-level variable $m$ that is used in place for a concrete address space.

\paragraph{Allocating Memory}
Dynamic memory allocations, i.e., allocations on the CPU heap and in global GPU memory, are managed via unique smart pointers to ensure that they are freed safely and without leaking memory.
We call the types of these values @-types, as they carry an annotation \emph{at} which address space they have been allocated.
The memory is freed when the smart pointer is destroyed at the end of a scope.
Therefore, our type $\code{T~@~cpu.mem}$ corresponds to $\code{Box\text{<}T\text{>}}$ in Rust and $\code{std::unique\_ptr\text{<}T\text{>}}$ in C++.
The following code shows how memory is allocated and initialised:
\begin{DescendListing}
{ let cpu_array: [i32,n] @ cpu.mem = CpuHeap::new([0;n]);
  { let global_array: [i32;n] @ gpu.global
          = GpuGlobal::alloc_copy(&cpu_array);
  } // free global_array
} // free heap_array
\end{DescendListing}
In the outer block, heap memory is allocated and initialised with an integer array of size $\code{n}$ filled with $0$.
The smart pointer that manages the allocation is then stored in variable $\code{cpu\_array}$.
In the inner block, GPU global memory is allocated for the data pointed to by $\code{heap\_array}$, the data is copied to the GPU and the resulting smart pointer is stored in $\code{global\_array}$.
The type annotations shown here are optional, but show the information stored in the type.

\subsection{Making Implicit Assumptions Explicit in \Descend{}}
\label{subsubsec:exec-gpu-program}
The CPU program is responsible for scheduling a GPU function for execution.
In \Descend{}, this happens with a special function call, as in CUDA, where not just the function arguments are provided, but also the executing GPU grid is specified; here comprising 32 block with 32 threads each:
\begin{DescendListing}
scale_vec::<<<X<32>, X<32>>>>(&uniq vec);
\end{DescendListing}
In contrast to CUDA, in \Descend{}, the GPU function signature carries the information what grid configuration is allowed to exectue the function:
\vspace{-1em}
\begin{DescendListing}
fn scale_vec(vec: &uniq gpu.global [i32; 1024])
    -[grid: gpu.grid<X<32>, X<32>>]-> ();
\end{DescendListing}

\Descend checks that the call side and the function declaration matches, to ensure that the assumptions about how the function is written and how it is invoked do not diverge.
\Descend{} also supports polymorphism over grid sizes, allowing GPU functions to be written that, for example, launch as many threads as the size of the input array.
In this case, the call side specifies the concrete values that are used for instantiating the grid size variables.

The CPU thread waits for the GPU function call to finish, meaning there is an implicit synchronization of the GPU grid at the end of each GPU computation.

\section{The Type System of \Descend{}}
\label{sec:descend-types}
In this section, we present the formal foundations of \Descend{}, including the formal syntax of terms and types as well as the most important typing rules, explaining the formal reasoning behind ensuring safety.
Our type system is based on the formalization of Rust's type system in Oxide~\cite{DBLP:journals/corr/abs-1903-00982}.
A technical report with the full type system of \Descend{} will be available at time of publication.

\subsection{Syntax of Terms}
{%
\floatstyle{boxed}
\restylefloat{figure}
\begin{figure*}
\begin{tabular}{l | l}
\begin{tabular}[t]{ l c l >{\em}r }
\tentry{ t }{ \sdef }{}{ Term: }
\tentry{}{}{ p }{ place expression }
\tentry{}{}{ \code{let}~x: \delta = t }{ definition }
\tentry{}{}{ p = t }{ assigment }
\tentry{}{}{ \&\optional{\Uniq}~p }{ (unique) borrow }
\tentry{}{}{ \code{\{}~\overline{t}~\code{\}} }{ block }
\end{tabular}
  &
\begin{tabular}[t]{ l >{\em}r }
  ${ \app{f}{\overline{\eta}, \overline{\mu}, \overline{\delta}}{\overline{t}} }$ & {\color{gray} function application }\\
  ${ \foreachloop{x}{t}{t} }$ & {\color{gray} for-each loop}\\
  ${ \fornat{n}{r_n}{t} }$ & {\color{gray} for-nat loop }\\
  ${ \Sched }$ & {\color{gray} schedule computation }\\
  ${ \indep{\optional{X \opt Y \opt Z}}{\eta}{e}{x}{t}{x}{t} }$ & {\color{gray} split execution resource }\\
  ${ \code{sync} }$ & {\color{gray} barrier synchronization }
\end{tabular}
\end{tabular}
\caption{Formal syntax of \Descend{} terms}
  \label{fig:syn-terms}
\end{figure*}
}
\Cref{fig:syn-terms} shows the formal syntax of terms of \Descend{}.
Place expressions are terms that express memory accesses, as discussed in \cref{sec:place-expressions}.
Let-bindings introduce and initialize new variables.
Assignments evaluate a term on the right of the equals sign and store the value in the memory referred to by the place expression on the left.
References are optinally annotated to be unique supporting writing through the reference.
By default references are read-only.
A block introduces a new scope consisting of a sequence of terms.
Function applications instantiate a polymorphic function $f$ with statically evaluated natural numbers ($\eta$), memories and data types, and call the resulting function with a list of terms as arguments.
There exist two kinds of for loops: a for-each loop over collections and a for loop over a statically evaluated range of natural numbers.
The scheduling primitive takes a dimension and schedules the same nested computation over the sub-execution resources nested within an execution resource, such as the threads in a block.
The split execution primitive splits an execution resource into two independent parts along the given dimension at the provided position.
It then specifies the computation each part performs within its body.
Finally, the barrier synchronization primitive synchronizes all threads within a block.

\subsection{Syntax of Types}
{%
\floatstyle{boxed}
\restylefloat{figure}
\begin{figure*}
  \centering
\begin{tabular}[t]{ l | r }
\begin{tabular}[t]{ l c l >{\em}r }
\tentry{}{}{}{}
\tentry{ \delta }{ \sdef }{}{ Data Types: }
\tentry{}{}{ \code{i32} \opt \ldots \opt \UnitTy }{ scalar type }
\tentry{}{}{ \TupleTy  }{ tuple type }
\tentry{}{}{ \ArrayTy \opt \llbracket \delta; \eta \rrbracket  }{array (view) type }
\tentry{}{}{ \&\optional{\Uniq}~\mu~\delta }{ reference type }
\tentry{}{}{ \delta~@~\mu }{ boxed-type }
\tentry{}{}{ x }{ type variable }
\tentry{}{}{}{}
\hline
\tentry{}{}{}{}
  \tentry{}{}{ <\overline{x:\kappa}>\TupleTy \xrightarrow{x:\varepsilon} \delta }{ Function Types }
\tentry{}{}{}{}
\hline
\tentry{}{}{}{}
\tentry{ \kappa }{ \sdef }{ \code{dt} \opt \Nat \opt \Mem }{Kinds}
\end{tabular}
  &
\begin{tabular}[t]{ l c l >{\em}r }
\tentry{ \eta }{ \sdef }{ 0 \opt \ldots \opt 9 \opt n \opt \eta + \eta \opt \eta * \eta \opt \ldots}{Nats}

\tentry{}{}{}{}
\hline
\tentry{}{}{}{}

\tentry { \mu }{ \sdef }{}{ Memory: }
\tentry {}{}
        { \CpuMem }{}
\tentry {}{}
        { \GpuGlobal }{}
\tentry {}{}
        { \GpuShared }{}
\tentry {}{}
        { m }{}

\tentry{}{}{}{}
\hline
\tentry{}{}{}{}

\tentry { \varepsilon }{ \sdef }{}{ Exec-Levels: }
\tentry {}{}
        { \CpuThreadTy }{}
\tentry {}{}
        { \gridty{d}{d} }{}
\tentry {}{}
        { \blockty{d} }{}
\tentry {}{}
        { \GpuThreadTy }{}
\end{tabular}
\end{tabular}

  \caption{Formal syntax of kinds and types in \Descend.}
  \label{fig:syn-types}
\end{figure*}
}
\Cref{fig:syn-types} shows the formal syntax of types and kinds.
Types in \Descend{} consist of data types ($\delta$) and function types.
Data types contain the standard scalar and tuple types.
Arrays are indexed by their size which is tracked symbolically in the type.
We introduce a special array view type for arrays that are transformed by views.
While standard arrays are guaranteed to be consecutive in memory, this is not the case for arrays with an array view type.
Reference types are modelled similarly to Oxide.
The $\Uniq$ qualifier marks a reference as unique.
Without a qualifier, references are shared---and read-only.
We extend the original definition with a memory annotation $\mu$ tracking the memory space the reference points to.
The possible memory address spaces are show on the right in the figure: $\CpuMem$, $\GpuGlobal$, and $\GpuShared$.
We omit here the presentation of lifetime variables that each reference carries, which are important for borrow checking, but complicate the presentation.
The treatment of lifetimes has been formalized in Oxide~\cite{DBLP:journals/corr/abs-1903-00982} and FR~\cite{DBLP:journals/toplas/Pearce21}.
Boxed-types track which memory space their smartly allocated value is stored in.
Finally, type variables can appear in polymorphic function definitions.

Function types can be polymorphic with a list of type-level variables, each annotated with their kind ($\kappa$).
Type-level variables can range either over data types, natural numbers ($\eta$), or memory spaces.
We currently restrict the function arguments and return types to be data types, ruling out higer-order functions in \Descend.
While it is easy to support this feature on the CPU it is not straightforward to implement higher order functions in an efficient way on the GPU and we, therefore, leave this for future work.
Above the function arrow, there is an additional parameter, the execution resource, which is annotated with an execution level ($\varepsilon$).
The execution level determines what execution resources are allowed to call the function by comparing the execution levels of the execution resource at the call side with the function annotation.

\subsection{Typing Rules}
\newcommand{\typejudge}[9]{%
  \def\tempa{#1}%
  \def\tempb{#2}%
  \def\tempc{#3}%
  \def\tempd{#4}%
  \def\tempe{#5}%
  \def\tempf{#6}%
  \def\tempg{#7}%
  \def\temph{#8}%
  \def\tempi{#9}%
  \typejudgecontinued
}
\newcommand{\typejudgecontinued}[2]{%
  \tempa\,;\,\tempb\,;\,\tempc\,;\,\tempd~~|~~\tempe\,;\,\tempf~~|~~\tempg~\vdash \boxed{\temph ~: \tempi}~\dashv~#1~~|~~#2
}
\newcommand{\pljudge}[7]{%
  #1\,;\,#2~~|~~#3\,;\,#4~~|~~#5~\vdash_{\code{pl}} #6 ~: #7
} 
\newcommand{\accjudge}[1]{%
  \def\tempa{#1}%
  \accjudgecontinued
}
\newcommand{\accjudgecontinued}[9]{%
  \set{#9} = \code{access\_safety\_check}(#8,#7,~~\tempa\,;\,#1\,;\,#2,#3\,;\,#4,#5)
}
\begin{figure*}

\begin{spacing}{1.3}
\begin{mathpar}
  \inferrule*[left=T-Sched]
  {
    \typejudge{\Delta}{\Gamma_g}{\Gamma_l}{\Theta}{e_f:\epsilon}{e\code{.forall(}d{\code{)}}}{\mathrm{A}}{t}{\UnitTy}{\Gamma_l'}{\mathrm{A}'}
  }
  {
    \typejudge{\Delta}{\Gamma_g}{\Gamma_l}{\Theta}{e_f:\epsilon}{e}{\mathrm{A}}{\sched{d}{x}{e}{t}}{\UnitTy}{\Gamma_l'}{\mathrm{A'}}
  }

  \inferrule*[left=T-Read-By-Copy]
  {
    \accjudge{\Delta}{\Gamma_l}{\Theta}{e_f:\epsilon}{e}{\mathrm{A}}{\overline{\pi}}{\Shrd}{p}{\loan{\Shrd}{p_i}}\\\\
    \pljudge{\Delta}{\Gamma_l}{e_f:\epsilon}{e}{\Shrd}{p}{\delta}\\
    \code{is\_copyable}(\delta)
  }
  {
    \typejudge{\Delta}{\Gamma_g}{\Gamma_l}{\Theta}{e_f:\epsilon}{e}{\mathrm{A}}
      {p}{\delta}
    {\Gamma_l}{\mathrm{A},e \rightarrow \mathrm{A}(e) \cup \set{\loan{\Shrd}{p_i}}}
  }

  \inferrule*[left=T-Write]
  {
    \accjudge{\Delta}{\Gamma_l'}{\Theta}{e_f:\epsilon}{e}{\mathrm{A}'}{\overline{\pi}}{\Uniq}{p}{\loan{\Uniq}{p}}\\\\
    \pljudge{\Delta}{\Gamma_l'}{e_f:\epsilon}{e}{\Uniq}{p}{\delta}\\
    \typejudge{\Delta}{\Gamma_g}{\Gamma_l}{\Theta}{e_f:\epsilon}{e}{\mathrm{A}}
      { t }{ \delta }
    {\Gamma_l'}{\mathrm{A}'}
  }
  {
    \typejudge{\Delta}{\Gamma_g}{\Gamma_l}{\Theta}{e_f:\epsilon}{e}{\mathrm{A}}
      {p~=~t}{\UnitTy}
    {\Gamma_l'}{\mathrm{A'},e \rightarrow \mathrm{A'}(e) \cup \set{\loan{\Uniq}{p_i}}}
  }

\end{mathpar}
\end{spacing}
  \caption{Important typing rules in \Descend for accessing memory for reading and writing, performing advanced access safety checks, as well as for scheduling computations over the execution hierarchy.}
  \label{fig:typing}
\end{figure*}

\paragraph{Typing judgement}
All safety checks in \Descend, including the crucial borrowing check, is formalized in the typing rules.
Therefore, the formal typing judgement is fairly involved with multiple environments, written as uppercase greek letters, to track various kinds of information.
The typing judgement considers information about the kinds of type variables ($\Delta$), the types of globally accessible functions ($\Gamma_g$), the types of local variables inside functions and active borrows ($\Gamma_l$), temporary borrows ($\Theta$), the execution resource executing the current function and its level ($e_f: \epsilon$), the execution resource executing the current statement ($e$), as well as the access environment $A$ that tracks which execcution recourse has access to which place expression.
Furthermore, the typing judgement is flow-sensitive, meaning that the typing and access environments change during the typing process.
For example when accessing an owned value we are not allowed to access it again (as it has been moved) and, therefore, it is removed from the typing environment.
Similarly, accesses changes from typing one expression to the next, as we will see below.

Therefore, the final typing judgement looks like
\begin{equation*}
     \typejudge{\Delta}{\Gamma_g}{\Gamma_l}{\Theta}{e_f:\epsilon}{e}{\mathrm{A}}
      {t}{\delta}
    {\Gamma_l'}{\mathrm{A'}}
\end{equation*}
which is saying, that term $t$ has type $\delta$ under the mentioned environments and produces the updated typing environment $\Gamma_l'$ and access environment $A'$.

\paragraph{Typing Rules}
We focus on three important typing rules in \cref{fig:typing}.
They give a more formal overview of what is required for one of \Descend's most important features: avoiding data races.
The rules are based on Oxide and adjusted to the additional requirements of GPUs with our execution resources and extended place expression syntax.

The \textsc{T-Sched} rule demonstrates how execution resources are tracked when navigating the execution hierarchy using $\code{sched}$.
The body of $\code{sched}$ is typed using the current execution resource $e$ extended by $\code{forall}$ with the appropriate dimension, to indicate that the body is executed by all execution resources in $e$.

The \textsc{T-Read-By-Copy} rule checks a place expression $p$ that is used to read a value from memory that is copyable, in contrast of values that are moved due to Rust ownership rules.
The rule states that $p$ is well-typed with data type $\delta$, if the premisses above the line are true.
To perform the check, a separate place expression typing judgment investigates the structure of $p$ to determine whether $p$ has type $\delta$.
This judgement also requires knowledge of whether $p$ is used in a shared or unique way, i.e., whether it is read or written, or whether it is borrowed sharedly or uniquely.
This is required to make sure that dereferences, of for example unique pointers, are permitted (which they are not if $p$ is used in a shared way).
The function $\code{is\_copyable}$ checks that the data type is copyable and not movable.
Finally, $\code{access\_safety\_check}$ performs the most crucial GPU-specific checks consisting of three logical steps:
\begin{enumerate}
  \item \emph{Narrowing check}: to check if the place expression is used uniquely by multiple execution resources.
  This check ensures that each execution resource selects its own distinct part from it.
  \item \emph{Access conflict check}: to check that using a place expression in an execution resource does not conflict with previous accesses by other execution resources that were stored in the access mapping environment
  \item \emph{Borrow checking}: performs the unchanged borrow checking as in Rust and as formalized in Oxide.
\end{enumerate}
For this rule, the check succeds if $p$ is sharedly accessible in the given environments.
The function computes a set of all possible aliases $p_i$ which are all marked as $\Shrd$.
The rule produces an updated access environment on the right side of the judgement in the conclusion.
We record, that the current execution resource $e$ accesses the place expressions $p_i$ in a shared way by adding the mapping to $A$.

Similarly to the prior rule, \textsc{T-Write} checks that the assignment of a term $t$ to place expression $p$ is safe, producing new typing and access mapping environments.
Term $t$ and place expression $p$ are both typed independently and must have the same type $\delta$.
As opposed to the previous typing rule, the typing judgment for place expressions and the access safety check are given the $\Uniq$ specifier to check that the write to memory is safe.
After type checking, the mapping in the access environment $\mathrm{A}$ is updated to reflect that the memory referred to by $p$ and its aliases was accessed uniquely.

A technical report detailing the entire type system of \Descend{} will be available at the time of the publication.

\section{Code Generation and Evaluation}
\label{sec:eval}
In this section, we give a brief overview of \Descend{}'s code generation implementation and evaluate \Descend code in comparison to handwritten CUDA code.
We will show that we can translate a program written in \Descend's holistic programming model into a CUDA program using the kernel programming model, without sacrificing performance.

\textbf{Code Generation}
The \Descend compiler translates \Descend code into CUDA C++ code.
\Descend functions for CPU threads are translated into C++ functions.
Functions that are run on the GPU are translated into CUDA kernels.
Before generating code, we inline function calls for functions whose execution resources are not a full grid on the GPU or thread on the CPU, such as functions executed by GPU blocks.

In CUDA's kernel programming model, all blocks and threads work concurrently.
This is exactly what the nested schedule primitives in \Descend are expressing.
Therefore, \lstinline!sched! does not appear in generated CUDA code, except for a scope that is introduced for its body and, the bound execution resource variable is compiled into the equivalent index identifying the thread or block in CUDA.
Block and thread indices are used when translating selections over place expressions into the raw memory index.
When selecting from or indexing into a view, these indices are transformed to express the access patterns these views describe.
This process is performed in reversed order, starting with the view that was applied last.
Each view takes the previous index and transforms it until the resulting index expresses a combination of all views.
The remaining \Descend syntax is translated straightforwardly, dropping static information that is not required in CUDA C++, such as memory annotations on reference types.

\textbf{Experimental Setup}
We performed an experimental evaluation using a Google Cloud instance with Debian GNU/Linux 10 and CUDA 11.6 on a Tesla P100 GPU.
We compare four different benchmarks that are commonly implemented for GPUs: block-wide parallel reduction, matrix transposition, scan and matrix multiplication.
Each algorithm was implemented in \Descend from which we generated CUDA code.
For the comparison, we implemented handwritten versions of the algorithms in CUDA using the same optimizations and access patterns.
All experiments were run for three different memory sizes: small, medium and large.
Using 256MB, 512, MB and 1GB of GPU memory.
We ran each benchmark 100 times, and measured the kernel runtimes.
The scan benchmark uses two different kernels and we measured the runtime from the start of the first until the end of the second kernel.

\begin{figure}
  \includegraphics[width=.48\textwidth]{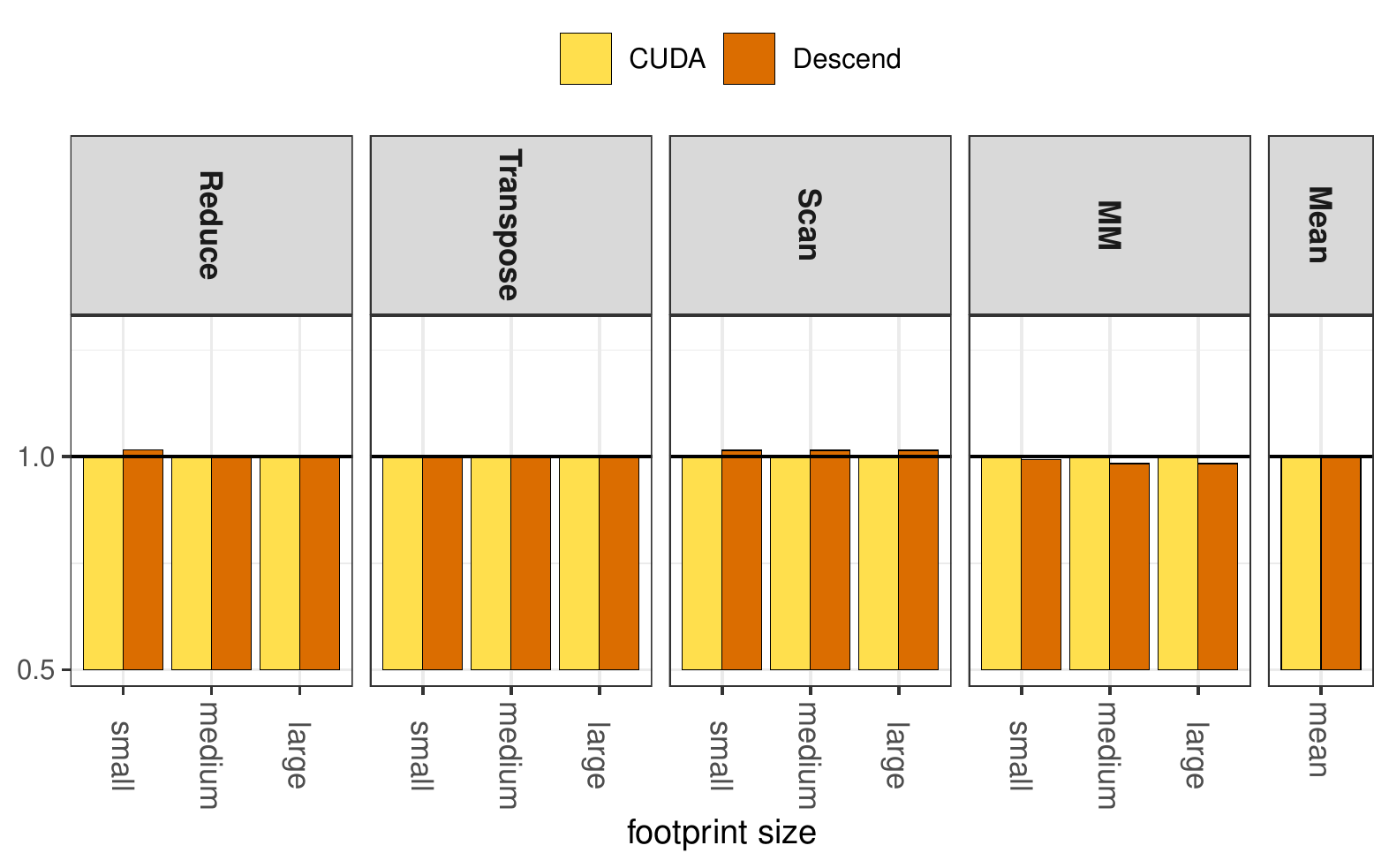}
  \caption{Relative runtimes between handwritten CUDA and Descend implementations. A higher bar indicates better performance.}
  \label{fig:measures}
\end{figure}

\textbf{Experimental Results}
\Cref{fig:measures} shows the relative median runtimes of \Descend compared to handwritten CUDA code.
It shows that \Descend and CUDA perform equally well for all benchmarks and sizes with performance difference of less than $3\%$.
We see that \Descend is expressive enough to write programs that achieve performance on-par with the handwritten CUDA implementations, while providing strong safety guarantees and catching bugs as demonstrated in \cref{sec:gpu-programming-difficult}.

\section{Related Work}
\label{sec:relatedWork}

\textbf{Unsafe GPU Programming Systems}
CUDA~\cite{DBLP:journals/queue/NickollsBGS08} is most likely the most popular GPU programming language.
OpenCL~\cite{opencl} and more recently SYCL~\cite{sycl} are vendor independent languages that follow a very similar design.
Many language bindings for languages other than C/C++ have been build, such as PyCUDA/PyOpenCL~\cite{DBLP:journals/pc/KlocknerPLCIF12} in Python, but they usually expose the CUDA programming model unchanged.
\cite{DBLP:conf/ipps/HolkPCLM13} extends Rust with the capability for expressing GPU programs and compiling them to PTX code.
GPU programs recognize shared and unique pointers, but are implemented in the traditional CUDA kernel programming model, maintaining all the problems identified in \cref{sec:gpu-programming-difficult}.

\textbf{Safe GPU Programming Systems}
There is a group of array languages with the goal of providing safe abstractions for high-performance GPU programming following functional ideas, including Futhark~\cite{DBLP:conf/pldi/HenriksenSEHO17}, Lift~\cite{DBLP:conf/cgo/SteuwerRD17} and its spiritual successor Rise~\cite{DBLP:journals/corr/abs-2201-03611}, and Accelerate~\cite{DBLP:conf/icfp/McDonellCKL13}.
In these languages, programs are safe by construction. 
They make use of functional patterns such as \emph{map} and \emph{reduce} to describe computations at a high level from which they generate low-level GPU code.
However, the high level abstractions come at a cost of loosing control, \Descend aims to empower programmers to exercise control with a safety net, as Rust promises it for the CPU.

\textbf{GPU Verification Tools}
One of the main goals of \Descend is to avoid data races.
There is previous work on static data race detection tools for GPUs like GPUVerify~\cite{DBLP:conf/oopsla/BettsCDQT12} and Faial~\cite{DBLP:conf/cav/CogumbreiroLRZ21}.
These tools analyze CUDA C code attempting to detect data races.
Faial creates a history of memory accesses, differentiating between read and write accesses.
This history is similar in spirit to the access mapping environment in \Descend's type system.
Of course the analyzed code may still contain a number of other problems that we mentioned in this paper and \Descend is capable to prevent statically.

\textbf{Formalizations of Rust}
Rust's ownerhsip, borrowing and lifetimes have been formalized in the FR Langauge~\cite{DBLP:journals/toplas/Pearce21} and Oxide~\cite{DBLP:journals/corr/abs-1903-00982}.
FR focuses on the core ideas of borrowing and lifetimes, and the way they are implemented for current Rust versions, while maintaining a maximally simple language with possible extensions.
Some practical features of Rust are not modelled in FR.
Oxide, which was a major basis for our work, focuses on borrowing rules of a new version of the borrow checker that is currently in development.
Furthermore, it formalizes other language features of Rust, such as polymorphic functions, slices, which can be seen as an early inspiration for \Descend's views and loops which are required for most practical applications.
Like Rust, these languages are not able to target GPUs.

\section{Conclusion}
\label{sec:conclusion}
GPU programming is notoriously challenging, but with \Descend, we have demonstrated that we can achieve the same performance as CUDA code while guaranteeing memory safety and statically rejecting programs with data races and incorrect synchronizations.
\Descend{} also assist programmers in managing CPU and GPU memory and enforcing previously implicit assumptions about the parallel execution of GPU code.
\Descend extends Rust's formal type system with \emph{execution resource} and \emph{views} to manage the GPU execution hierarchy and ensure safe parallel memory accesses.

\bibliographystyle{plain}
\bibliography{bib}

\end{document}